# Numerical study of boiling of Liquid Nitrogen on a liquid-liquid contact plane

Rupak Kumar, Arup Kumar Das

Department of Mechanical and industrial engineering, Indian Institute of Technology, Roorkee, 247667, India

## Abstract

In this paper, a numerical study is conducted to investigate boiling of a cryogen on a solid surface as well as on a liquid surface. Both single mode and multi-mode boiling is reported for boiling on to a solid surface. In case of boiling on a liquid surface, liquid nitrogen ($LN_2$) is selected as the cryogen (boiling fluid) and water is chosen as the base fluid (heating fluid). Different flow instabilities and their underlying consequences during boiling of a cryogen are also discussed. For the boiling on a solid surface, in the single mode, bubble growth, its departure, and area weighted average heat flux are reported, where they increase linearly with increase in the wall superheat. Asymmetry in the bubble growth and departure of $2^{nd}$ batch of the vapor bubbles have been observed due to local fluctuations and turbulence created just after the pinch off of the $1^{st}$ batch of vapor bubbles in case of multi-mode boiling on the solid surface. Boiling of $LN_2$ on a liquid surface is reported for a base fluid (Water) temperature of 300 K. Vapor film thickness decreases with time and the minimum film thickness just before rupture is 7.62 $\mu m$, dominance of thermocapillary over vapor thrust causes breaking of the vapor film at t = 0.0325 s. The difference in evaporation rate and vapor generation, before and after vapor film collapse is significant.

## Introduction

Boiling of a liquid (boiling fluid) in direct contact with another liquid (heating fluid) can be found in many industrial applications as well as in several natural incidents. One such scenario occurs during accidental spills of liquefied natural gases (LNG) while transportation via cargos in sea water. LNG, which is composed mostly of methane and other higher alkanes such as ethane, propane, butane and nitrogen, when comes in direct contact with water may transform its phase rapidly depending upon the local flow conditions and several other factors. This rapid transformation of phase also known as rapid phase transition (RPT) generally occurs whenever there is absence of any imperfections on the surface, it can be a mirror finished surface or a liquid-liquid contact plane. Boiling fluid gets superheated beyond its saturation temperature ($T_S$), causing it to change its phase homogeneously at an instant. Rapid phase transfer and thereafter change of state, causes gas to expand quickly which resembles like a physical explosion. This violent explosion may initiate shock waves and can harm equipment and people in the vicinity, also, if the gas is flammable and comes in contact with a fire source, it can cause some serious damage. Therefore, there is a need to fully understand this complex phenomena of liquid-liquid boiling and the critical factors which causes this sudden change of phase and methods to control this behavior.

Superheat limit of a fluid is the maximum attainable temperature at a given pressure. Liquids upon reaching this superheat limit changes its phase from liquid to vapor phase instantly which also expands simultaneously just after the phase change to ambient pressure which looks like a physical explosion. Hence, in the past several attempts by different authors were made to calculate the superheat limit of the various fluids. Blander and Hengstenberg (1971) calculated the homogeneous nucleation temperature for different fluids and mixtures using steady state Zeldovich-Kagan theory and verified the results experimentally. They found that the theory predicts accurately the superheat limit for pure liquids but insisted on requirement of further investigation for binary mixtures. Rausch and Levine (1973) developed a thermodynamic and heat transfer model to predict the superheat limit of the cryogen to cause a shock wave, which they found out to be 0.84 times the critical temperature ($T_C$) of the cryogen. Based on which they also provided an expression to predict the necessary temperature of the base fluid to superheat the cryogen up to its superheat limit ($T_{SL}$) and eventually cause a shock wave. They verified the expressions with the experimental results of different fluids. They concluded that if the cryogen is a zeotropic mixture, the selective boil-off of the lighter hydrocarbon will make the mixture rich in the heavier hydrocarbons, which in turn will increase the mixture critical temperature, and hence the superheat limit and thus shock will be strongly dependent upon the composition of the mixture. Reid (1976) experimentally investigated superheating of a liquid droplet (n-pentane) using a bubble column filled with sulfuric acid and wrapped with a heating element, wrapped closely at the top compared to the bottom so that the top is warmer compared to the bottom, and a liquid drop is injected at the bottom which explodes the moment it reaches the warmer region where the local temperature is equal to its superheat limit. In this study, based on the kinetic analysis, he found that the liquids must be superheated in order to form vapor bubbles in the bulk fluid, since the small vapor bubbles are quite unstable and collapses readily. Nishigaki and Saji (1981) verified an empirical law with the experimental results for the superheat limit of oxygen and nitrogen. Sinha et al. (1987) experimentally determined the superheat limit of the liquid nitrogen using a transient heating method for different temperature and pressure range.

After understanding the superheating, the primary reason behind the rapid phase transition (RPT), and its quantification, authors have focused on physical and thermodynamic understanding of this complex behavior of RPT. Enger et al. (1973) proposed a physical model describing the mechanism of boiling of liquid-liquid systems like liquefied gas-water systems, where they claimed that the lack of nucleation sites shifts the transition region towards higher temperature, superheating the boiling fluid. They concluded that the superheating of the boiling fluid depends upon the temperature difference between the fluids and the composition of LNG, also, vapor explosion would not occur in case the methane content of LNG is less than 40 mol %. They also claimed that the energy released upon vapor explosion of liquefied-gas water systems is limited by the superheat limit of the fluid. Drake et al. (1975) conducted and experimental study to compare the boil-off rates of different cryogenic fluids, liquid nitrogen ($LN_2$), methane and ethane upon boiling on a water surface. They found that all the three cryogenic fluids behaved differently upon spillage onto water surface, $LN_2$ boiled in a film boiling regime with minimum heat flux but maximum vapor superheat, ethane boiled in nucleate boiling regime with maximum heat flux and minimum superheat, methane on the hand found to be the intermediate case, where both nucleate and film boiling regime occur, heat flux and vapor superheat ranges between the data obtained for $LN_2$ and ethane.

Vapor film collapse between the base fluid and the cryogenic fluid leads to superheating of the cryogenic fluid and later shifting of boiling regime from film boiling to transition and finally nucleate boiling. Yalencia-Chavez (1973) studied the confined spillage of LNG on water surface and the effect of its composition on the evaporation rates. Experiments were performed to evaluate evaporation rates for pure methane, ethane, propane, their mixtures and LNG. He observed that methane initially boils in the film boiling regime, until the formation of an ice layer which promotes nucleate boiling, ethane boils in transition boiling regime and again formation of ice shifts the regime to nucleate boiling, but propane boils in nucleate boiling regime since beginning, whereas LNG mixtures evaporates in a preferential manner, where, more volatile component evaporates first, this causes rise in the saturation temperature of the residual liquid. LNG mixture starts to boil with film boiling regime but preferential evaporation causes methane to evaporate quickly, decreasing the vapor pressure in the vicinity of base of the bubble, which causes vapor film collapse. He concluded that, with the increase in amount of higher hydrocarbons in the LNG, probability of vapor film collapse also increases. Dincer at al. (1976) experimentally investigated the effect of initial temperature of water on evaporation rate of liquid nitrogen ($LN_2$) and methane on water. They concluded that if the initial temperature of the water is low, the energy transfer from water surface to the cryogenic fluid is provided by latent heat of formation of ice at the surface with little effect on underlying water, whereas at higher temperature of water surface, energy is supplied through convection with homogeneous cooling of water, formation of ice surface also leads to shift in boiling mechanism from film to nucleate boiling, which was also observed by Yalencia-Chavez (1973). Vesovic (2006) studied the effect of ice formation on the evaporation of LNG over water body. He also concluded from this study that formation of ice leads to sharp change in temperature of water surface leading to vapor film collapse and change of mechanism from film to nucleate boiling.

Vapor dispersion after the rapid phase transition is major cause of concern, hence, several studies have been conducted to predict the vapor dispersion, and various ways to control the dispersion. Benjamin et al. (2008) identified the key parameters which affect the vapor dispersion and used a CFD model to find their effect on the vapor dispersion upon spillage of LNG. Studies were conducted for two cases, in the first case LNG was released on a water surface and in the other case on a concrete surface. Some of the key parameters which had high impact on the vapor dispersion for the first case i.e. LNG released on water body, and in the presence of high as well as low wind velocity includes release rate, wind velocity and sensible heat flux from the ground, whereas some key parameters had high impact only at low wind velocity such as pool area, velocity profile of the gas phase and the turbulence at the source, while other parameters such as pool shape and geometry, ambient temperature, surface roughness and wind direction have relatively low impact as compared to other parameters discussed before. A similar study was also performed by Qi et al. (2010) where they also discussed the key parameters such as evaporation rate, pool area, atmospheric conditions, turbulence in the source term, ground surface temperature, height of the roughness and obstacles. They compared the numerical result with the experimental data and found to be in good agreement. Recently, Horvat (2018) developed a CFD based model to simulate spillage of LNG on water surface, its spreading, rapid phase transition and finally vapor dispersion. They discussed the effect of temperature of vapor cloud on its dispersion, initially just after the phase change, the temperature of vapor cloud at the saturation temperature i.e. -162 ℃, and hence it is negatively buoyant and vapor clouds floats just above the water surface, but after mixing with the surrounding air, when its temperature goes above -108 ℃, it become positively buoyant which lifts the vapor cloud from the water surface and helps in its dispersion. They observed that the effect of RPT event on vapor dispersion is limited but RPT has significant effect on the shock wave formed due to sudden expansion of vapor cloud. Some studied have also been conducted to contain the vapor cloud dispersion.  Rana et al. (2010) conducted field experiments to understand the vapor cloud dispersion and ways to mitigate the dispersion upon LNG spillage on the land. They used two different techniques, water spray curtains and flat fans to control the vapor cloud dispersion, where they found water spray curtains provide insignificant dilution of vapor cloud. However, the flat fans

Experimental studies have been conducted by numerous authors to model the source term in order to describe this complex phenomena of cryogenic boiling through numerical study. One such study was conducted by Vechot et al. (2013) where they investigated the effect of different heat transfer mechanism i.e. conduction, convection and radiation, and their individual contribution on the vaporization of cryogenic fluid, liquid nitrogen ($LN_2$). They used a Dewar flask to to store the $LN_2$ also minimize heat transfer through conduction and an electronic weighing machine to measure the vaporization rate. They conducted two sets of experiments to measure the vaporization rate of $LN_2$, in the first set, they divided the experiment into four groups, where in the first group they checked the effect of only conduction through walls and the lid at the top, then in the second group, along with the conduction, effect of convection in the room and natural radiation, then in the third group, effect of radiation from a bulb (200 W), convection in the room along with the conduction heat loss from all sides, and finally effect of forced convection where they used an electric fan, along with the natural radiation in the room and conduction through walls, in the second set of experiments, effect of forced convection are investigated, where they varied the wind speed from 1.9 m/s to 3.3 m/s, and also considered the conduction from all the sides. They observed that the vaporization rate increased significantly in case of forced convection and ration heat transfer, but contribution from convective heat transfer was significantly higher compared to other heat transfer mechanisms. Gopalaswami et al. (2015) experimentally investigated the vaporization heat flux of liquid nitrogen ($LN_2$) continuously released onto water surface, where they used an electronic balance to measure the mass flux of vapor based onto the mass loss data for different spill rates. They predicted the heat flux using the energy balance for the nitrogen-water system. They also developed a boiling regime map for liquid nitrogen boiling on water using the experimental results, they compared the experimental results with the predicted results based on the expression developed by Berenson (1961) for film boiling heat transfer on a horizontal surface, which they found to be predominantly in the film boiling regime. They found the vaporization flux to be independent of amount of $LN_2$ spilled, however the initial flux was dependent on the water temperature and spill rate. They also observed that the amount of vapor generation increases with spill area and the size of water body.

Morse and Kytomaa investigated the effect of turbulence on evaporation rate of liquid nitrogen ($LN_2$) and LNG. They calculated the turbulence intensity by measuring the temporal variation of the interfacial height at $LN_2$-water interface. The velocity magnitude variations were measured by comparing the potential and kinetic energy at the interface. They found that the rate of evaporation increases considerably with turbulence intensity for the spillage of liquid nitrogen ($LN_2$) on water as compared to that of LNG. Stutz et al. (2013) experimentally investigated the role of evaporation waves in the propagation of boiling phenomenon in a superheated n-pentane on a copper block. They have used a mirror finished surface to reach the superheat limit. Thy concluded that in the case of a very high degree of liquid superheat, boiling on a mirror finished surface doesn't follow the classical bubble growth geometry, in its place, a single bubble is seen first followed by phase change around that bubble which spreads around that bubble in all directions, which was referred as "Straw hat structure". Hence, they proposed a physical model based on the theory of evaporation waves which explains the phase change phenomena for this case. The proposed model divided the whole phenomena into three different physical phenomena, where at first the vaporization front was visualized as the evaporation wave, then the superheated liquid was partially vaporized in the vaporization front while remaining form small droplets and impinge on the surface which depending upon the heat flux go through partial phase change, and at last the development of straw hat structure drives the superheated liquid and changes the flow field nearby.

Very few authors have worked on the numerical study of boiling of a cryogenic fluid, one such study was conducted by Liu et al. (2015) where they developed a model for cryogenic boiling on a solid surface, model includes both heterogeneous and homogeneous mode of boiling. The model successfully predicted different boiling regimes, nucleate, transition and film boiling for different wall superheats. They developed a boiling regime map for liquid nitrogen ($LN_2$). They also investigated the effect of liquid level on the heat flux, which decreases with the increase in the liquid level. In a similar study, Ahammad et al. (2016 a&b) developed a CFD model for film boiling of a cryogenic fluid on a solid surface. They discussed the effect of wall superheat onto the average heat flux, bubble frequency and bubble diameter, they also proposed model based on the energy balance to predict the heat flux in the case of film boiling. Both the studies discussed were based on Rayleigh-Taylor instability (RTI), and are discussed in detail in the results and discussions sections.

Considering the literature review, a very limited studies have been conducted for boiling of a cryogen on the solid surface, moreover all the studies have been performed for single-mode boiling assuming steady and uniform behavior throughout the domain. A multi-mode boiling for the cryogenic fluid on a solid surface has never been conducted before, moreover to the best of knowledge of the authors, interfacial study of boiling of a cryogenic fluid on a liquid surface has also not been performed before. Hence, in the present study, boiling of a cryogenic fluid ($LN_2$) has been investigated on a solid (single-mode and multi-mode) as well as on a liquid surface.

## 2. Mathematical model

Following section describes the computational domain, mesh structure, governing equations, phase change model and surface tension model utilized for this study.

### *2.1. Computational domain, mesh structure and boundary conditions*

A computational domain of width of $\lambda_{d/2}$ and a length of $3\lambda_{d/2}$ (Figure 1) is chosen for the study of boiling of $LN_2$ on solid surface, where $\lambda_{d/2}$ is the most dangerous wavelength (Carey, 1992), given by equation (1).

$$\lambda_d = 2\pi \left[\frac{3\sigma}{(\rho_l - \rho_v)g}\right]^{1/2} \quad (1)$$

Where, $\sigma$ is the interfacial tension between liquid nitrogen ($LN_2$) an nitrogen vapor ($N_2$), and $\rho_l$, $\rho_v$ are density of $LN_2$ and $N_2$ respectively. A uniform structured mesh of 64 x 192 segments is selected for the present study based on grid independence test. Pressure outlet boundary condition is set up at the top, while symmetry boundary condition on either adjacent walls, at the bottom wall, no-slip boundary condition is set. Both liquid nitrogen ($LN_2$) and nitrogen vapor ($N_2$) are initially set at 77 K. The initial vapor film (Figure 1) is patched as a sinusoidal disturbance based on the Rayleigh-Taylor instability and given by equation (2), also. A linear temperature gradient is also applied in the initial vapor film (Ahammad et al., 2016) and given by equation (3).

$$\delta = \frac{\lambda_d}{64}\left(4 + \cos\left(\frac{2\pi x}{\lambda_d}\right)\right) \quad (2)$$

$$T_y = \begin{cases} T_W - \Delta T \cdot \dfrac{y}{\delta}, & \alpha = 1 \\ T_{sat}, & \alpha = 0 \end{cases} \quad (3)$$

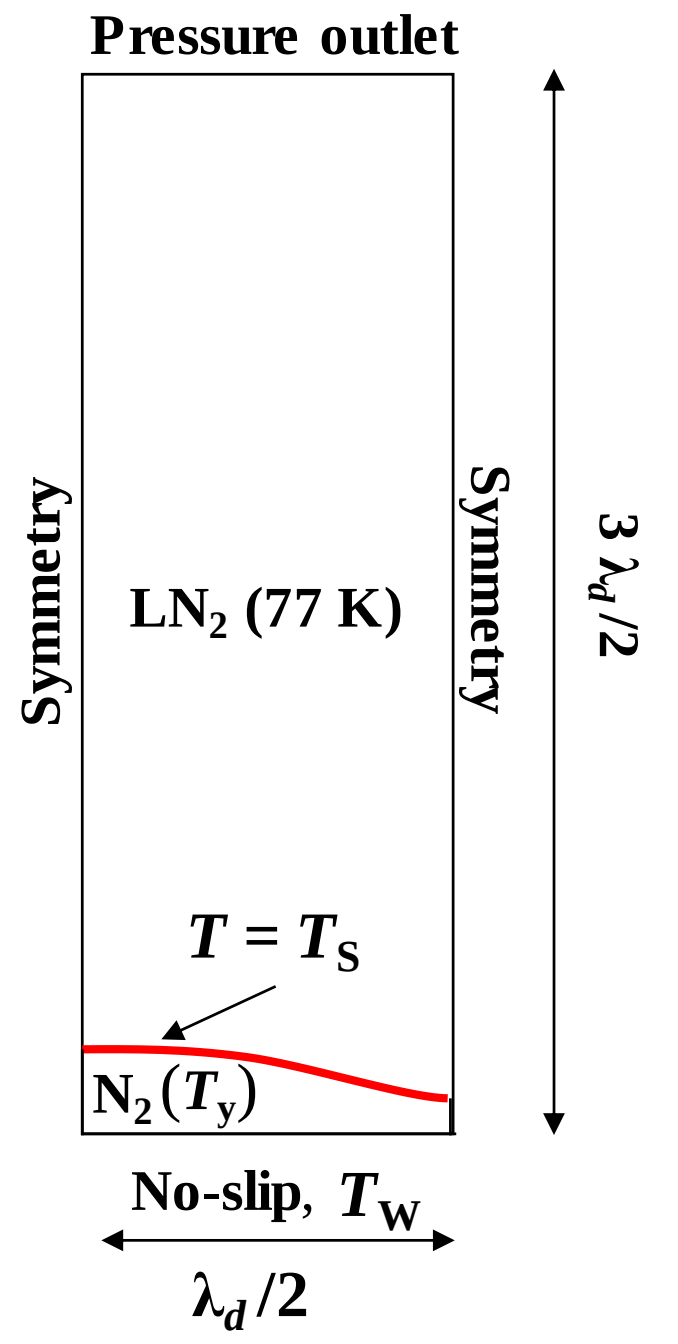

**Figure 1.** Schematic of computational domain along with the boundary conditions

*2.2. Governing equations and computational setup*

The equation for mass conservation (4), momemtum conservation (5), energy conservation (8), phase fraction transport equation (9), phase change model (13,14 & 15) and surface tension model (16) are as follows:

*Continuity equation*

$$\frac{\partial \rho}{\partial t} + \boldsymbol{\nabla} \cdot (\rho\, \boldsymbol{v}) = S_m \tag{4}$$

Continuity equation is by equation (4), where, $\rho$ and $\boldsymbol{v}$ are density and velocity vector respectively, whereas $S_m$ is the source term to account for the mass transfer phase during boiling.

*Momentum equation*

$$\frac{\partial}{\partial t}(\rho\, \boldsymbol{v}) + \boldsymbol{\nabla} \cdot (\rho\, \boldsymbol{vv}) = -\boldsymbol{\nabla} p + \boldsymbol{\nabla} \cdot \{\mu(\boldsymbol{\nabla}\, v + \boldsymbol{\nabla} v^T)\} + \rho \boldsymbol{g} + \boldsymbol{F_{st}} \tag{5}$$

$$F_{st} = \sum_{i<j} \sigma_{ij} \frac{\alpha_i \rho_i \kappa_j \boldsymbol{\nabla} \alpha_j + \alpha_j \rho_j \kappa_i \boldsymbol{\nabla} \alpha_i}{\frac{1}{2}(\rho_i + \rho_j)} \tag{6}$$

$$\kappa_i = \frac{\nabla \alpha_i}{|\nabla \alpha_i|}, \kappa_j = \frac{\nabla \alpha_j}{|\nabla \alpha_j|} \tag{7}$$

Momentum equation is given by equation (5) where, $p$, $\mu$, $g$ are pressure field, dynamic viscosity and gravitational constant respectively, whereas $F_{st}$ is the surface tension force given by equation (6). It is based on the continuum surface force model, given by Brackbill et al. (1992). In equation (6), $\sigma_{ij}$, $\kappa_i$ and $\kappa_j$ are interfacial surface tension between phase *i* and phase *j* , curvature of phase *i* and *j* respectively. The curvatures of phase *i* and phase *j* are given by equation (7).

*Energy equation*

$$\frac{\partial(\rho E)}{\partial t} + \nabla \cdot \big(\boldsymbol{v}(\rho E + p)\big) = \nabla \cdot (k_{\text{eff}} \nabla T) + S_e \tag{8}$$

Energy equation is given by equation (8) where $k_{\text{eff}}$ , $S_e$ are effective conductivity and volumetric heat source term to account for energy transfer during phase change respectively. Source term $S_e$ is given by equation (15).

Transport equation for phase fraction

$$\frac{\partial \alpha_i}{\partial t} + \boldsymbol{\nabla} \cdot (\alpha_i\, \boldsymbol{v}) = 0 \tag{9}$$

$$\sum_i^{\mathcal{N}} \alpha_i = 1 \tag{10}$$

Interfaces between different fluids are tracked by solving transport equations (9) individually for each phase which is called volume of fluid (VOF) method given by Hirt and Nicholas (1988). Here, $\alpha_i$ is the phase fraction of $i^{th}$ phase. If $\alpha_i$ is zero, in that case the cell is empty of $i^{th}$ phase and if $\alpha_i$ is one, that cell is filled with $i^{th}$ phase alone and if $\alpha_i$ varies between zero and one, there is an interface between $i^{th}$ phase and other phases. Sum of all the phases in a given cell is always equal to one (equation 10).

$$\rho = \sum_{i=1}^{N} \rho_i \alpha_i \tag{11}$$

$$\mu = \sum_{i=1}^{N} \mu_i \alpha_i \tag{12}$$

Mixture properties used in continuity, momentum and energy equations is given by equation (11 & 12) which is phase fraction weighted linear sum of individual property of different phases.

*Phase change model*

Phase change model utilized in the study considers both heterogeneous and homogeneous mode of boiling (Liu et al., 2015). Equation (13) and equation (14) has been employed to account for heterogeneous and homogeneous boiling mode respectively.

$$S_{m2} = \frac{(k_l \alpha_l + k_v \alpha_v)(\nabla T \cdot \nabla \alpha_l)}{L} \tag{13}$$

$$S_{m1} = r \alpha_\ell \rho_\ell \left( \frac{T - T_S}{T_{sat}} \right) , T \geq T_S \tag{14}$$

$$S_e = -S_m . L \tag{15}$$

Here, $k$, $\alpha$, $L$, $r$, is thermal conductivity, phase fraction, latent heat of evaporation and evaporation frequency respectively where subscripts $'l'$ denotes liquid phase and $'v'$ denotes vapor phase.

*Surface tension model*

Surface tension ($\sigma$), has been utilized in the study so as to capture the thermocapillary effect, where $\sigma$ is linearly dependent upon the temperature given by equation (16), where, $\sigma_0$ is the surface tension at the saturation temperature and $\gamma$ is the surface tension gradient (equation (17).

$$\sigma(T) = \sigma_0 - \gamma(T - T_S) \tag{16}$$

$$\gamma = -\frac{\partial \sigma}{\partial T} \tag{17}$$

A finite volume based commercially available computational fluid dynamics software ANSYS Fluent is used to carry out the numerical simulation for the problem. All the fluids are assumed to be incompressible, Newtonian and immiscible. Volume of fluid method (VOF) is employed to track different interfaces. Pressure-Implicit with Splitting of Operators (PISO) scheme is used for pressure-velocity coupling. For spatial discretization, pressure staggering option (PRESTO) for pressure, least squares cell based for gradient and Geo-Reconstruct for volume fraction, second order upwind scheme for momentum is used. For transient discretization, a second order implicit scheme is used.

## 3. Results and discussions

An attempt is made to develop a model for boiling of liquefied Nitrogen ($LN_2$) which has a very low boiling point of 77 K at atmospheric pressure on a solid as well as liquid surface. Boiling over a liquid medium is quite different from boiling on a solid surface. One of the important parameters in case of boiling on a solid surface, imperfections, is absent for the case of boiling on a liquid surface. In the absence of imperfections, boiling phenomena which generally occurs at the saturation temperature may be delayed up to an upper limit called superheat limit ($T_{SL}$) of the liquid. Boiling on a liquid surface can happen anywhere between saturation temperature ($T_S$) and superheat limit depending upon local flow conditions, thermal properties and interfacial interactions. Such superheating of the boiling fluid may lead to instantaneous phase change which may appear explosive some time in nature. A proper understanding of this superheating phenomena and consequently explosive boiling and the critical parameters which affect this behavior is of paramount importance for the safety concerns. Hence, in the following sections, boiling of a cryogenic fluid on a solid surface and subsequently on a liquid surface has been discussed.

### *3.1. Boiling of a cryogen on a horizontal solid surface*

Numerical computation of boiling on a horizontal solid surface has been reported extensively before for various fluids such as water and others (Son and Dir, 1997; Tomar et al., 2005), but boiling of a cryogen has not been targeted by many. The very first attempt to investigate boiling of a cryogen on a horizontal surface was tried by Liu at al. (2015), where they have predicted the boiling nature of $LN_2$ for various wall superheats. They have also developed a model for boiling of a cryogenic liquid where two different bubble formation routes have been explained. The first route considers the vapor generation at the horizontal surface which occurs in the case of heterogeneous boiling where the vapor present at the initial stage (t = 0 s) continues to grow due to evaporation at liquid-vapor interface. The other route takes into account the vapor generation in the bulk cryogenic liquid which generally occurs in case of homogeneous boiling where vapor is generated due to superheating of the bulk fluid whenever it comes in direct contact with the horizontal solid surface.

The mass transfer models for the evaporation at the liquid-vapor interface (heterogeneous boiling) and evaporation in the bulk fluid (flash or homogeneous boiling) are expressed as:

$$S_{m1} = \frac{Q_1}{L} = \frac{-(k_l\alpha_l + k_v\alpha_v)\nabla T \cdot \nabla\alpha_l}{L} \tag{1}$$

$$S_{m2} = \frac{Q_2}{L} = \frac{-\alpha_l\rho_l C_{p,l}(T_l - T_s)}{L\, r} \tag{2}$$

where, $k$, $\alpha$, $L$, $C_p$ and $r$ denotes heat conductivity, volume fraction, latent heat of evaporation, specific heat and time constant, respectively. Simulations have been performed by Liu et al. (2015) for 8 different wall temperatures covering nucleate, transition and film boiling which further resulted in a boiling regime map for $LN_2$. Significant presence of small bubbles (micro bubbles) in case of nucleate boiling have been reported which later diminishes and later on converted into film at higher temperatures. They have also investigated the effect of depth of $LN_2$ and found that the heat flux generation is reduced with the increase in the level of $LN_2$.

A similar study was also performed by Ahammad et al. (2016a) where they have employed CFD model to investigate saturated film boiling of $LN_2$ as well as LNG, considering it as pure methane. Growth and departure patterns of vapor bubbles for various wall superheats for both i.e. $LN_2$ and LNG were discussed. In their subsequent work (Ahammad et al. 2016b), film boiling heat transfer is predicted in case of boiling of a cryogen on a horizontal solid surface for both $LN_2$ and LNG. In both the above studies, a Rayleigh-Taylor instability (RTI) based model (Zuber et al., 1959) for saturated film boiling on a horizontal solid surface have been employed. The most dangerous wavelength ($\lambda_d$), as selected by Liu et al. (2015) is $\sqrt{2}\,\lambda_c$ whereas the same has been taken as $\sqrt{3}\,\lambda_c$ by Ahammad et al. (2016). Both the studied were conducted for single mode boiling, and it was assumed that the behavior of the vapor bubbles will be repeated throughout the computational domain. No work has been reported to study multi-mode boiling heat transfer till date. In present work, boiling of a liquid nitrogen ($LN_2$) is simulated on a horizontal solid surface for both single and multi-mode occurrences. Role of different flow instabilities such as Rayleigh-Taylor, Benard-Marangoni and Rayleigh-Benard is discussed, in the context of boiling. Calculation of wall heat flux and Nusselt number in case of boiling on a horizontal solid surface is discussed, in details. Bubble growth behavior and its departure pattern is also reported for different degree of superheats. In Multi-mode boiling, asymmetricity in the bubble growth and its departure is also highlighted during departure of 2$^{nd}$ batch of vapor bubbles.

### *3.1.1. Flow Instabilities in boiling of a cryogen*

Flow instabilities play major role in boiling of a cryogenic fluid and hence, consideration of same in the development of a numerical model is essential. Two important hydrodynamic instabilities that have significant role in boiling are, Rayleigh-Taylor instability (RTI) and Kelvin-Helmholtz instability (KHI). Presence of a denser fluid over a lighter fluid causes instability along the interface which grows over time until the lighter fluid pinches off. This instability is observed during saturated film boiling and called Rayleigh-Taylor instability (RTI). As discussed before, based on the RTI, Zuber et al. (1959) was first to develop a model for saturated film boiling on a horizontal surface. He came up with two important wavelengths, most critical wavelength ($\lambda_c$), below which there will not be any bubble formation, most dangerous wavelength ($\lambda_d$), distance between two adjacent vapor bubbles for which the vapor generation will be maximum. In KHI, when a lighter fluid floats over a denser fluid and the fluids have different velocities, then instability is induced along the fluid-fluid interface which may grow in time, this can be observed in case of spillage and pool spreading of LNG over water surface, where LNG (lighter fluid) spreads over the water surface (denser fluid) upon spillage.

Apart from hydrodynamic instability, temperature induced flow instability also plays important role in case of film boiling on a horizontal surface. Two important thermal induced flow instabilities are Benard-Marangoni instability (BMI) and Rayleigh-Benard instability (RBI). In BMI, when a very high temperature is applied to a very thin film of capillary length scale, a surface tension gradient is induced along the interface which tries to pull the interface from the region of low surface tension to high surface tension. Thermocapillary instability has a major role in vapor film collapse during saturated film boiling of a cryogenic fluid on a horizontal solid surface (Aursand et al., 2018). Marangoni number (Ma) is used to characterize the BMI and given by $Ma = \frac{\gamma h \Delta T}{\nu k}$, where $\gamma$ is the surface tension gradient with respect to temperature, $h$ is the film thickness, $\Delta T$ is the temperature difference along the interface, $\nu$ is kinematic viscosity and $k$ is thermal conductivity. Unlike BMI, Rayleigh-Benard instability which occurs in the bulk fluid of considerable depth, here, a density gradient is induced along the depth of the bulk fluid induced by the temperature gradient. This instability causes a local Rayleigh-Benard convection in the region, as fluid particles with high temperature becomes less dense and tries to move up and replace the denser cold particles lying above. One must note that this instability doesn't play any role in the film boiling of a cryogen on solid surface rather it's important in the case of boiling on a liquid surface but discussed here for coherency. Rayleigh number characterizes the RBI, which is given by $Ra = \frac{\alpha g h^3 \Delta T}{\nu k}$. Here $\alpha$, $g$, $h$, $\Delta T$, $\nu$ and $k$ are thermal expansion coefficient, acceleration of gravity, film thickness, temperature difference in the region, kinematic viscosity and thermal conductivity, respectively.

### *3.1.2. Wall Heat flux*

In the context of random bubble formation in boiling over a surface, the area weighted average heat flux can be evaluated by:

$$\frac{1}{A}\int q dA = \frac{1}{A}\sum_{i=1}^{n} q_i |A_i| \tag{1}$$

where, $A$, $q$ , $n$ and $i$ are total area of the wall, heat flux associated with the wall, and total number of facets and individual facet, respectively. Also. Nusselt number (Nu), which is an important dimensionless number characterizing the boiling heat transfer with respect to wave length ($\lambda_0$) as length scale, can be calculated as:

$$Nu = \frac{|q"|\lambda_0}{k_l(T_{wall} - T_s)} \tag{2}$$

Figure 1. Shows the variation of area weighted wall heat flux derived from present simulation for a wall superheat of 103 K along with acceptable range of validation against the results of Ahammad et al. (2017). At t = 0.09 s and 0.12 s of first bubble cycle, film thickness is minimum and maximum, respectively, and subsequently these are the maximum and minimum values of heat flux and Nusselt number in Figure 1.

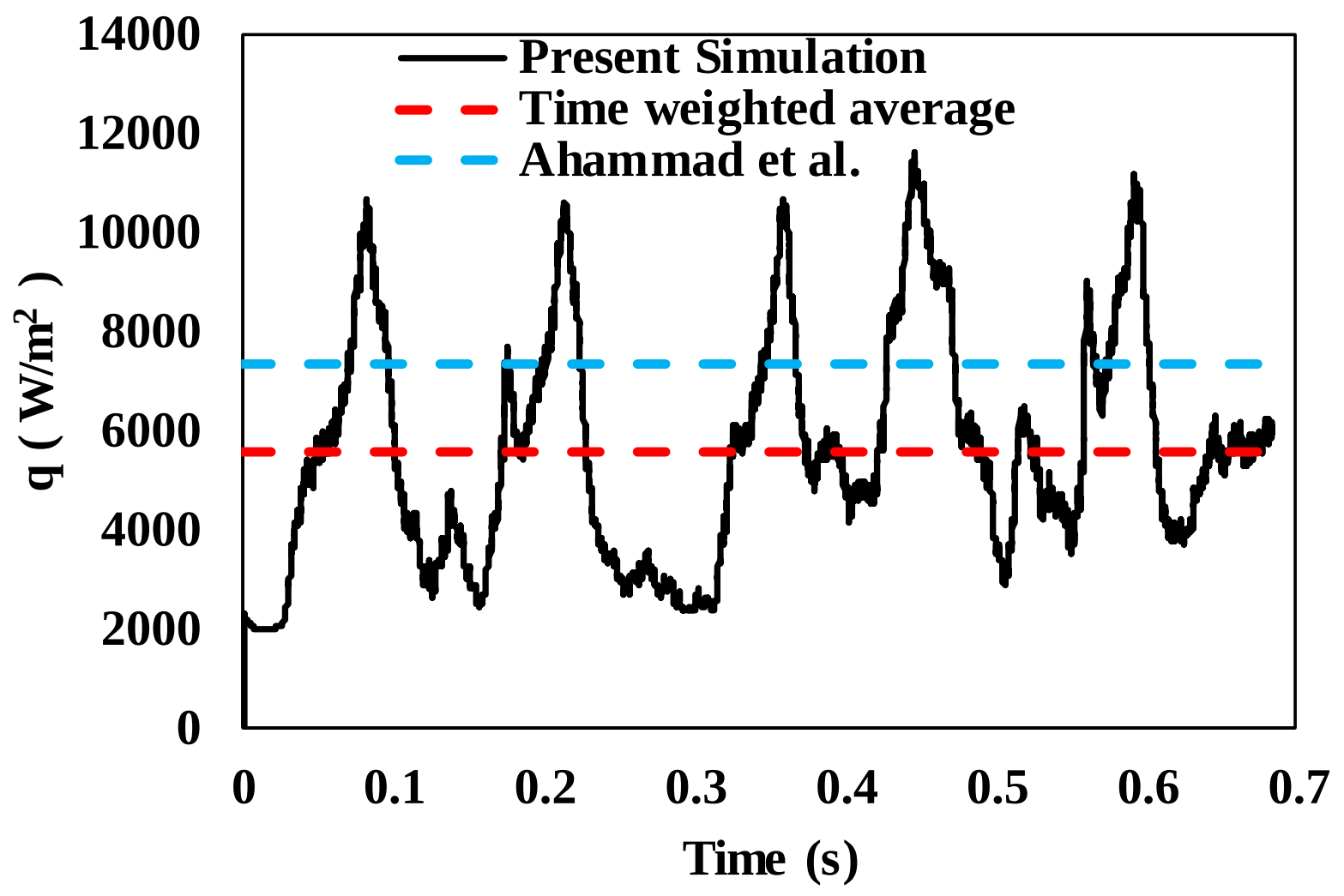

**Figure 1:** Wall heat flux variation for the present simulation as compared to Ahammad et al. (2017)

### *3.1.3. Bubble growth and departure*

Transient evolution of vapor bubble for single mode boiling is shown in Figure 2 for a wall superheat of 223 K. Initially at t = 0.05 s, the vapor film is stable as the gravitational force dominates over the buoyancy force. With the temporal growth of the initial perturbation of nitrogen vapor due to evaporation of $LN_2$ at the liquid-vapor interface as shown in Figure 1(a), nitrogen vapor moves towards the nodal point where the initial thickness is maximum. This movement of nitrogen vapor towards the nodal point helps in growth of initial disturbance as can be seen in Figure 1(b). Disturbance continues to grow as the heavier fluid $LN_2$ lies above the lighter fluid $N_2$, a classic case of Rayleigh-Taylor instability (RTI). As the vapor bubble grow further in size (Figure 1(c)), necking appears at the base of the vapor bubble, which eventually leads to the pinch-off of the vapor bubble (Figure 1(d)). As discussed before, the film thickness is minimum just before the departure of the vapor bubble which also corresponds to the maximum Nusselt number and heat transfer coefficient. Similar behavior was also reported by Ahammad et al. (2016) and Liu et al. (2015) in their studies.

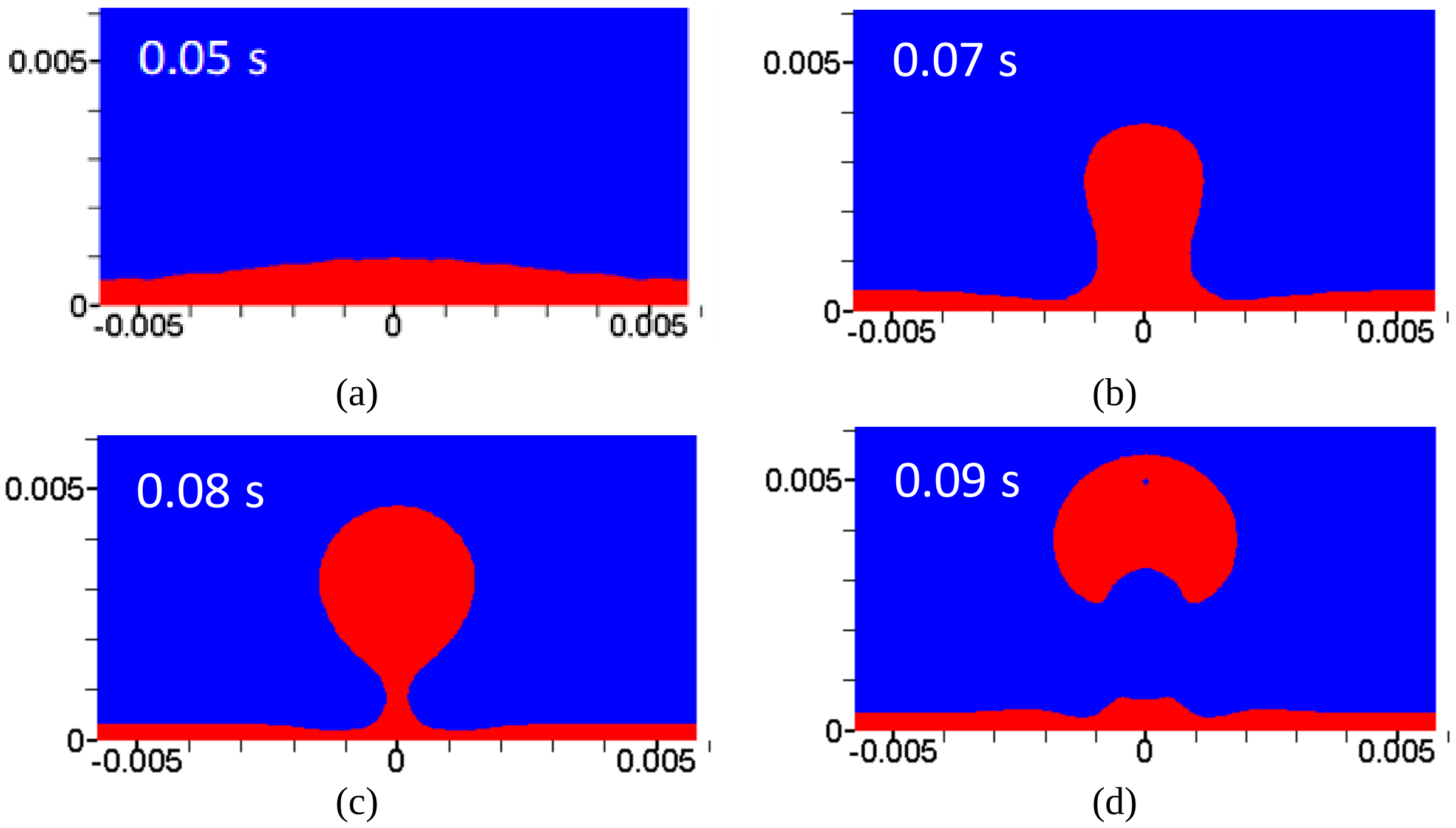

**Figure 2**: Growth of vapour bubble and its departure for a wall superheat of 223K.

### *3.1.4. Effect of wall superheat*

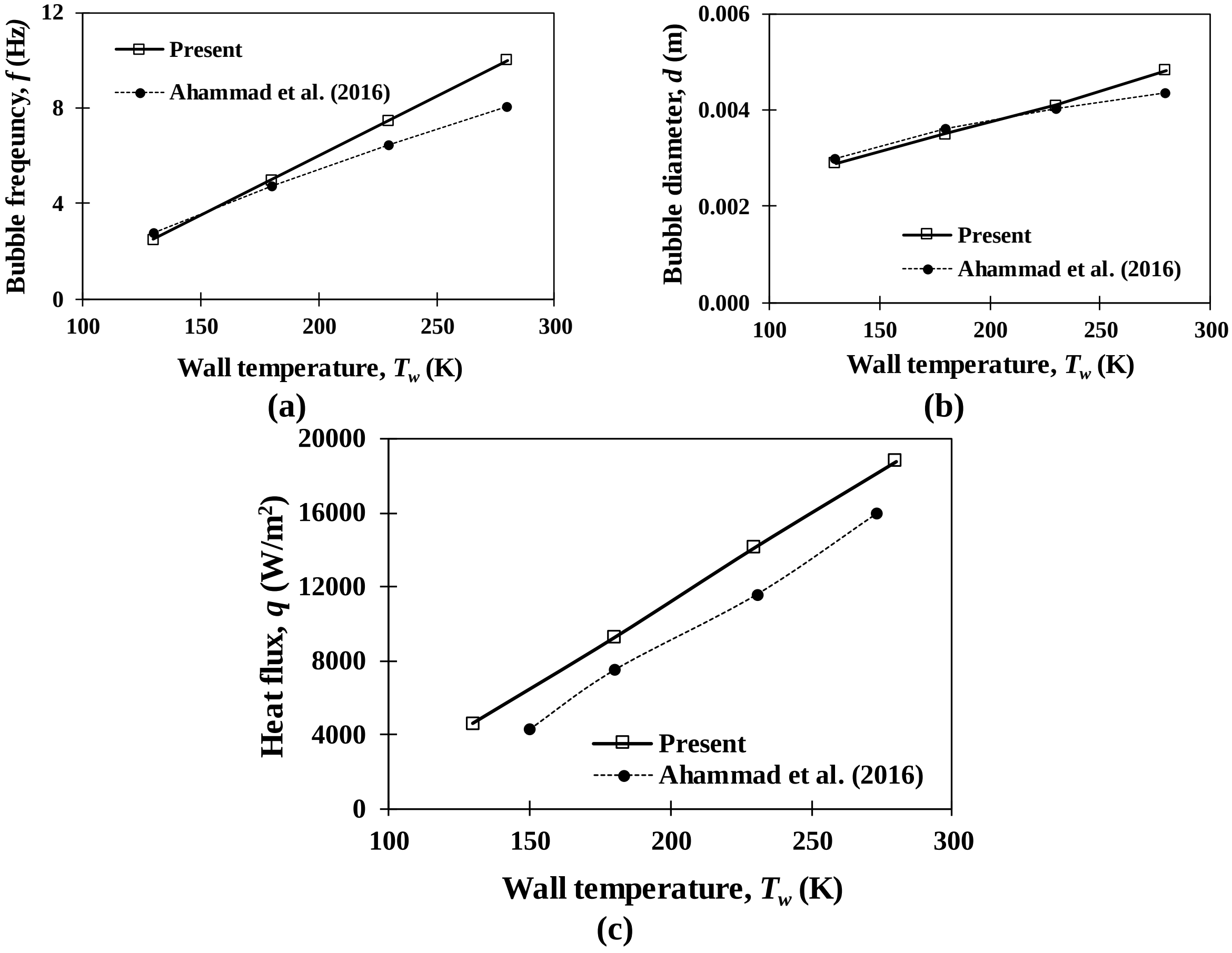

**Figure 3:** (a) Rate of bubble generation, (b) change in bubble diameter, and (c) heat flux generated due to phase transfer due to increase in wall temperature.

Simulations are performed for different wall superheats and its effect on vapor bubble frequency, bubble diameter and heat flux are reported (Figure 3). Both bubble frequency and bubble diameter increases linearly with the increase in the wall temperature (Figure 3(a & b)), since increasing the wall temperature increases the rate of evaporation at the liquid-vapor interface, which causes faster growth and departure of bubbles. Area weighted average heat flux, $q$, generated at the bottom wall also increases linearly with the increase in the wall temperature since the heat flux is directly dependent on the temperature difference between wall and the saturation temperature of boiling fluid. All the results, i.e. bubble frequency, bubble diameter and heat flux are compared against the results of Ahammad et al. (2016) and similar behavior were also reported by them.

#### *3.1.5. Multi-mode boiling and asymmetry in bubble departure*

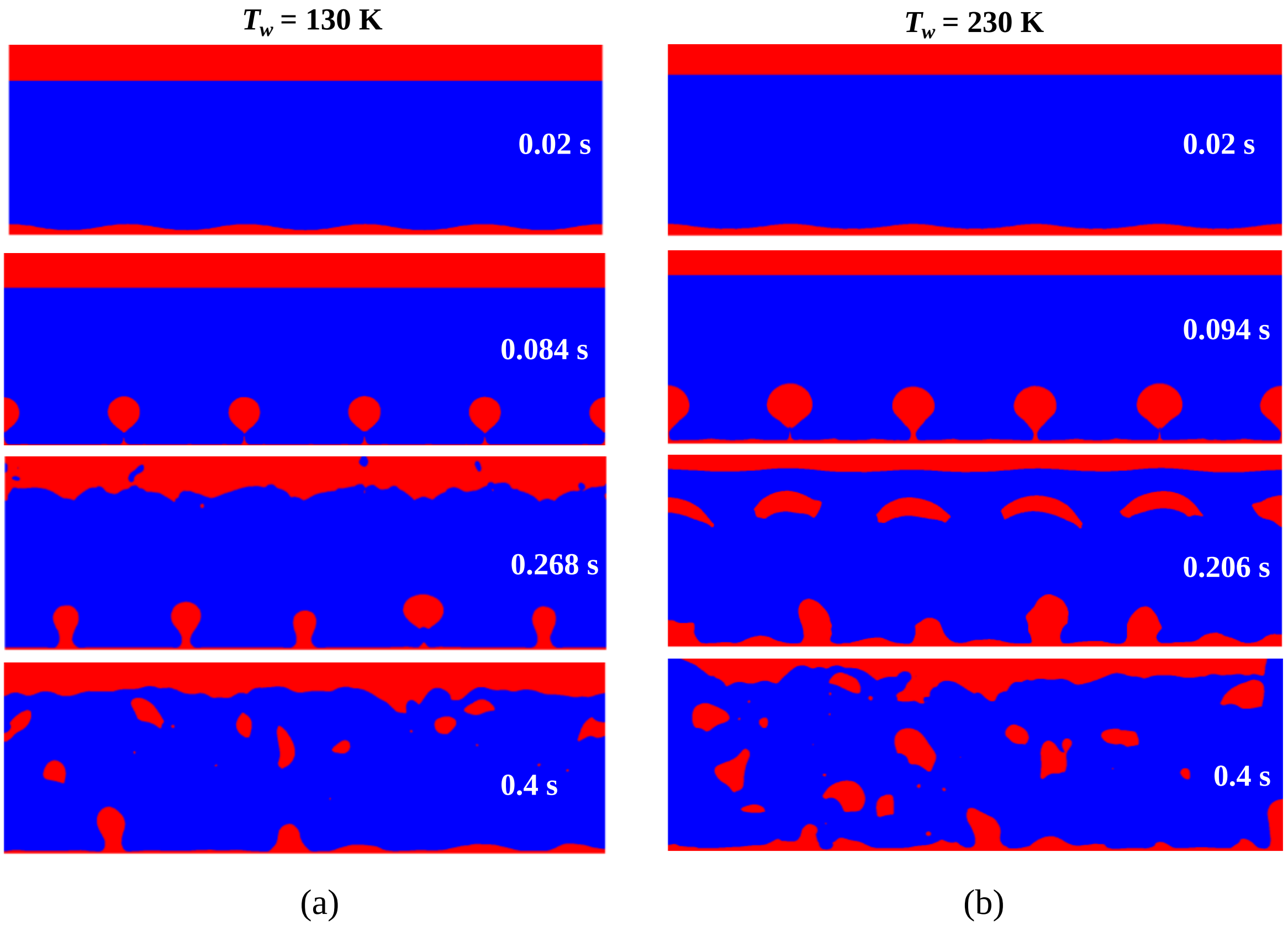

**Figure 4:** Growth and behaviour of vapour bubbles for multi-mode boiling on a horizontal solid surface for two different wall temperature (a) 130 k, and (b) 230 K

Most of the studies conducted (Liu at al., 2015; Ahammad et al., 2016) for boiling of a cryogenic fluid on a horizontal solid surface was in single mode. Present simulation, on the other hand, for the first time, simulates a multi-mode boiling of a cryogenic fluid on a horizontal solid surface. All the single mode studies assume that the behaviour of vapour bubbles will remain steady and uniform throughout, but that is not the case in experiments. Therefore, a multi-mode boiling case for the same conditions is simulated for a computational domain of size $5\lambda_{d/2}$ x $10\lambda_{d/2}$ mm$^2$, where $\lambda_d$ is the most dangerous wavelength (Zuber et al., 1959). Boundary conditions are same as before, i.e., pressure outlet at the top, no-slip at the bottom wall, and symmetry at the adjacent walls. A constant temperature thermal boundary condition is given at the bottom wall.

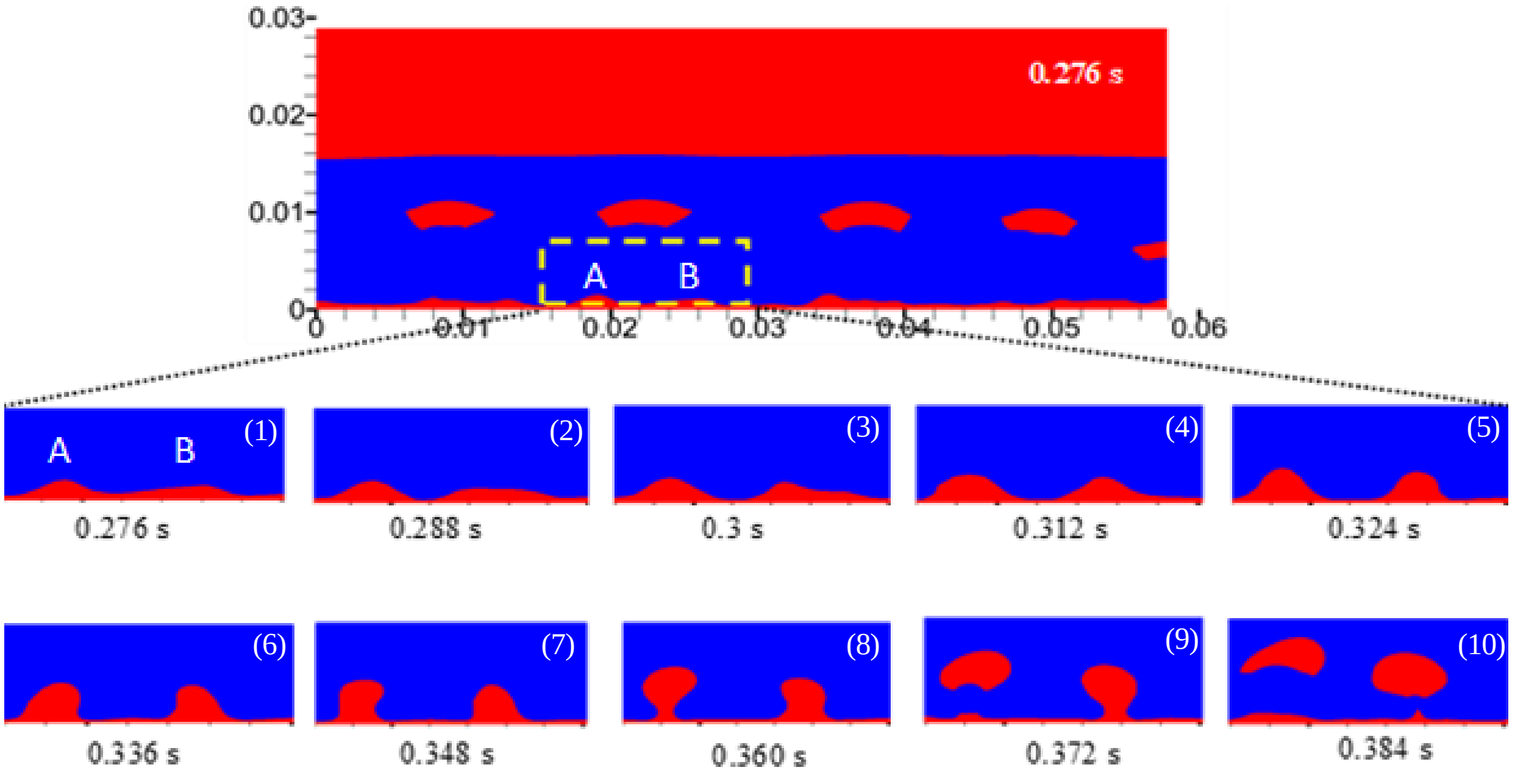

(a)

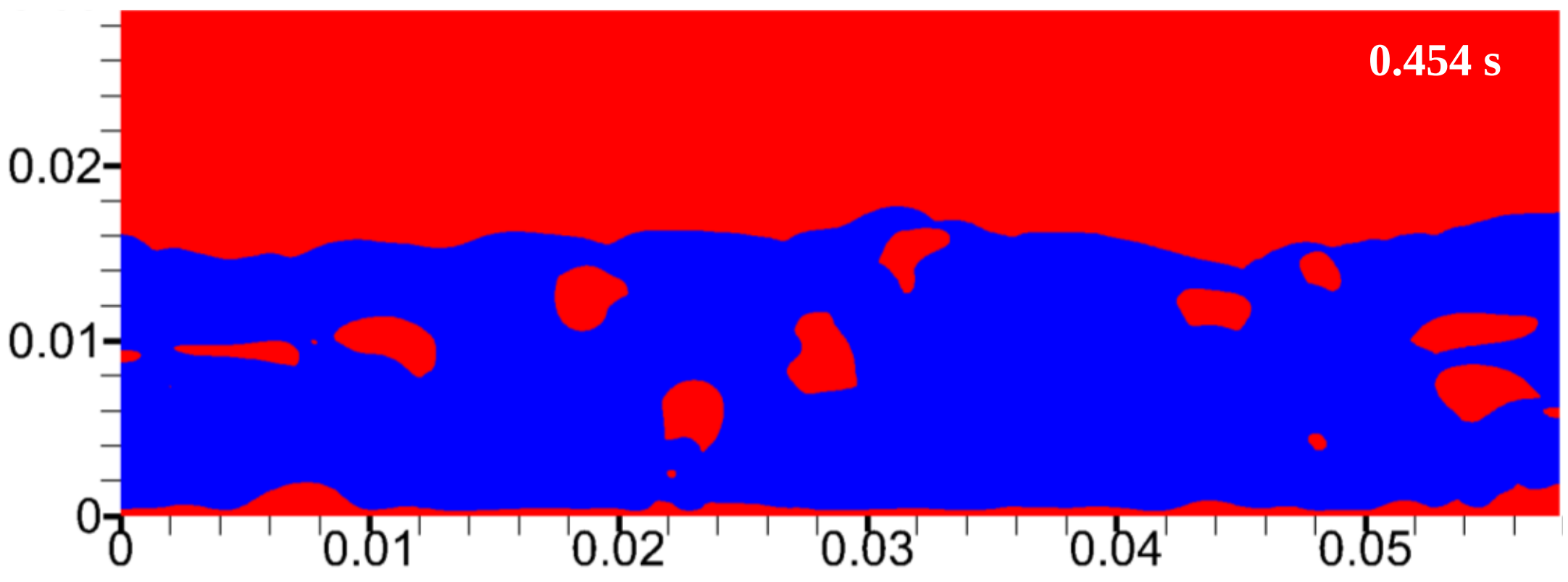

(b)

**Figure 4:** (a) Asymmetry in bubble growth and the departure of 2nd batch of vapour bubbles, and (b) chaotic behaviour before merging with the bulk fluid in multi-mode boiling, for a wall superheat of 223 K.

Growth and departure for the 1st batch of vapor bubbles and subsequent batches are shown in Figure 4 (a & b) for two different wall temperatures of 130 K and 230 K. As evident from the phase contours, the first batch of vapor bubble grow and depart in a similar fashion as compared to single mode without interfering with the behavior of their neighbors. The 2nd batch of vapor bubbles have differences in their sizes and departure time, which is more evident for higher wall superheat. After the departure of 2nd batch of vapor bubbles, it becomes more chaotic as can be seen in the Figure 4(a & b) just before the departure of 3rd batch of vapor bubbles.

The 2nd batch of vapor bubbles compete with their neighbours for growth and departure. Figure 4(a) shows the behaviour of two neighbouring bubble vapour sites for various time steps which covers their growth and asymmetry in departure. The wall superheat for this case is 223 K. Growth and departure of 1st batch of vapor bubbles occur in an idealized condition, where, each and every perturbation are of equal size, moreover the temperature distribution in the perturbations are also same. But, that is not the case for 2nd batch of vapor bubbles, where initial perturbations are not of the same size and have different amount of vapor content, and the temperature distribution is also different for every vapor sites.

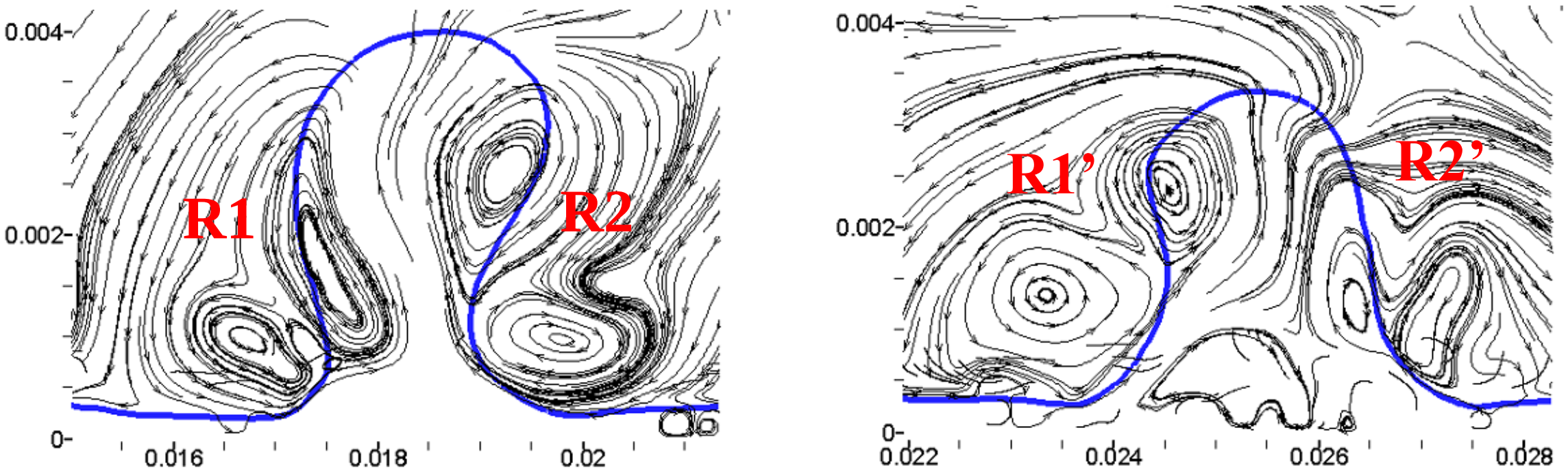

(a) (b)

**Figure 5:** Presence of turbulence fluctuations causing asymmetry in bubble departure

Apart from this, due to the pinch-off of the 1st batch of vapor bubbles, a local turbulence fluctuation in the flow field is created which causes asymmetry in the bubble growth for the 2nd batch of the vapor bubbles, which is also a reason of asymmetry in bubble growth and their departure (Ruth, 2019). Moreover, the temperature distribution in the film thickness after the departure of 1st batch of vapor bubbles is not uniform anymore which may have also caused uneven vapor generation along the liquid-vapor interface.

In the Figure 4(a), at t = 0.276 s, because of the local turbulence fluctuations created by the departure of 1st batch of vapor bubbles, shape of bubbles A and B are quite different, bubble B is more spread out in the form of a plateau whereas bubble A has conical shape, which can also be seen at t = 0.288 s. At t = 0.324 s, bubble B also takes the conical shape, but bubble A clearly is more voluminous. With the progression in time, as the bubbles become larger in size, they also become asymmetric in the latitudinal direction (t = 0.348 s) due to the turbulence fluctuations. Figure 5 (a & b) shows the streamlines near the bubbles A and B at t = 0.348 s. Notably, the vorticities near the base of bubble A (Figure 5(a)) is stronger as compared to that of bubble B (Figure 5(b)). The same can be attributed to the local turbulent fluctuations. If we consider only bubble A, uneven fluctuation across it in latitudinal direction, as indicated by region R1 and R2 in Figure 5(a), may have caused the asymmetry for the bubble A. similar behavior can also be observed for bubble B, as specified by region R1' and R2' in Figure 5(b). At t = 0.348 s, necking also appears early for bubble A, necking grows further and becomes millimetric in size at t = 0.360 s, difference in size of neck between them is quite large and finally at t = 0.372 s, bubble A pinches off before the bubble B which takes 0.012 seconds more to pinch off afterwards.

### 3.2. *Boiling of a cryogen on liquid surface*

Study of cryogenic fluid on a liquid surface is of paramount importance considering the variety of applications of concern such as cryogen spills on water. In such cases, it is crucial to avoid the collapse of stable vapor film, since collapse of vapor film leads to direct contact of boiling fluid and the heating fluid. In the absence of any bubble-forming nuclei, such contacts may lead to superheating of the boiling fluid. Superheating may cause rapid change of phase at an instant which appears explosive in nature, and if the vapor is flammable and there is also a presence of light source, it can be really detrimental for the safety concerns of general well-being. Hence, the boiling regime to be continuously in film boiling regime is desirous. So it is very important to identify the critical parameters which affect this collapse, also, in the case of any collapse in the vapor film, finding some innovative ways to either avoid this collapse or in the worst case cease the vapor diffusion.

Boiling of a cryogenic fluid on a solid surface differs from the boiling on a liquid surface, and these differences can be categorized into two parts, (a) fluid dynamics, and (b) thermodynamics. Considering the fluid dynamics first, the normal thrust of the vapor phase will be different for boiling on a solid than that on a liquid surface since the liquid surface is deformable in nature. Secondly, considering the thermodynamics, the rate of diffusion in case of solid will be faster as compared to that of liquid. Apart from this, in case of liquid-liquid boiling, due to the presence of temperature gradient in the bulk fluid, a density gradient will be induced which will cause a local Rayleigh-Benard convection few millimeters below the hot liquid surface (Kumar et al., 2020).

Boiling of a cryogen on a liquid surface is simulated here in single mode for a computational domain of width $\lambda_{d/2}$ and a length of $5\lambda_{d/2}$. The liquid surface chosen for the study is water, and $LN_2$ is the cryogen. Three phases in the study include water, $LN_2$ and $N_2$ vapor. Initially, water phase is filled up to a height of 15 mm from the bottom of the computational domain, over which a $N_2$ vapor is patched given by equation (2). Rest of the space is occupied by the $LN_2$ phase. Pressure outlet boundary condition is used on top, no-slip boundary condition at the bottom wall, symmetry at the adjacent walls, and all the walls are taken to be adiabatic. Initially, water phase and $LN_2$ are taken to be of uniform temperature of 300 K and 77 K respectively, whereas the temperature distribution in the $N_2$ vapour is given by:

$$T_y = \begin{Bmatrix} T_W - \Delta T \cdot \dfrac{y}{\delta}\ , \ y \le \delta \\ T_s \qquad\qquad , y > \delta \end{Bmatrix} \tag{17}$$

Figure 6 shows the transient evolution of the phase contours as obtained from simulation. At t = 0.021 s, initially patched vapor phase receives the heat diffused from the water surface which leads to evaporation of $LN_2$ to at the liquid-vapor interface. The vapor film thickness continuously reduces, as can be seen from the Figure 6, and at t = 0.320 s, it reaches its minimum value before collapse, and finally at t = 0.325 s, the vapor film collapses. After the breaking of vapor film at t = 0.0325 s, the cryogen $LN_2$ comes into direct contact with the liquid surface, water, and in the absence of bubble-forming nuclei, causes the superheating of $LN_2$, up to an upper limit called superheat limit ($T_{SL}$).

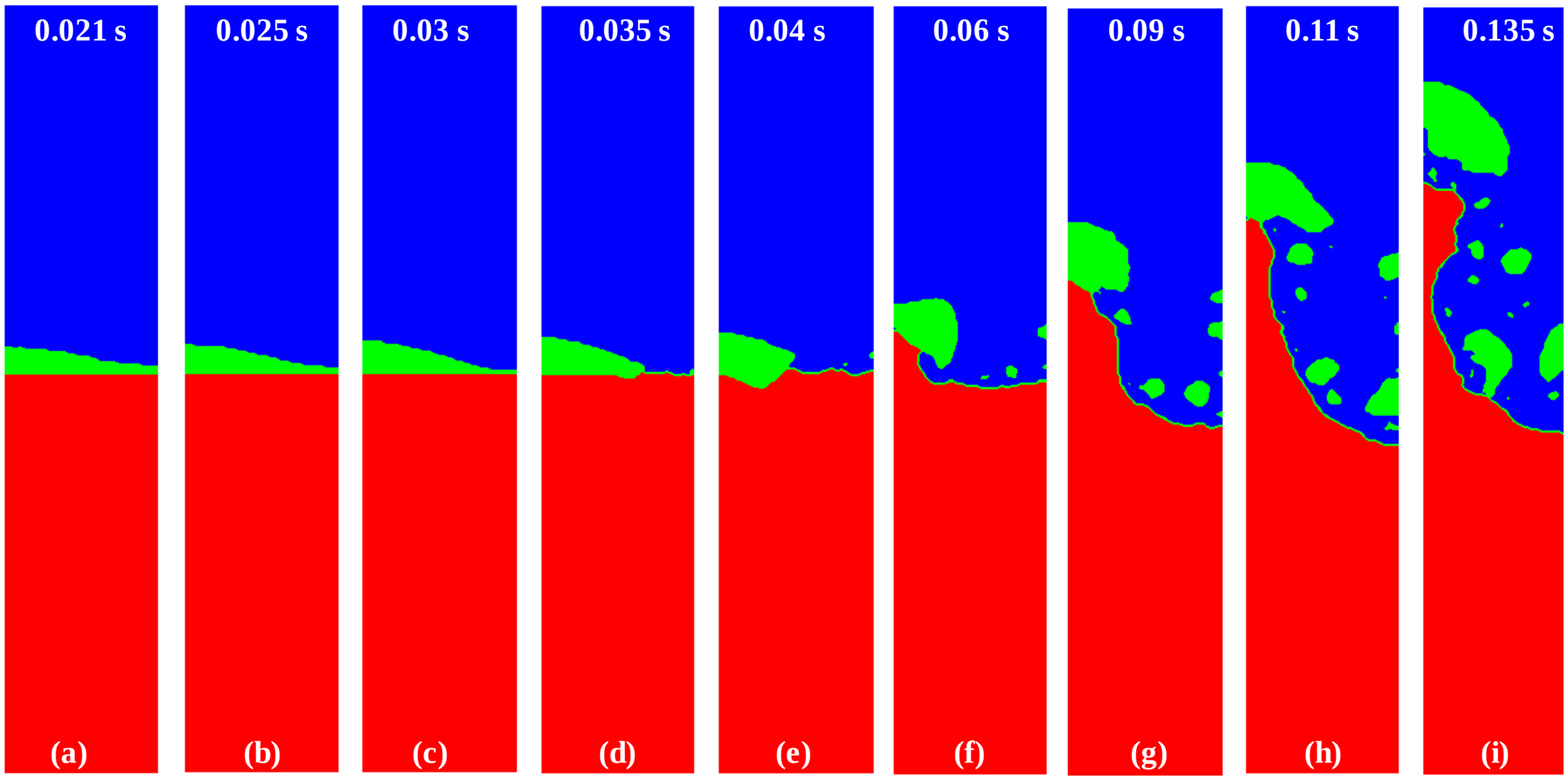

Figure 6: Transient evolution of phase contours of boiling of $LN_2$ on water surface

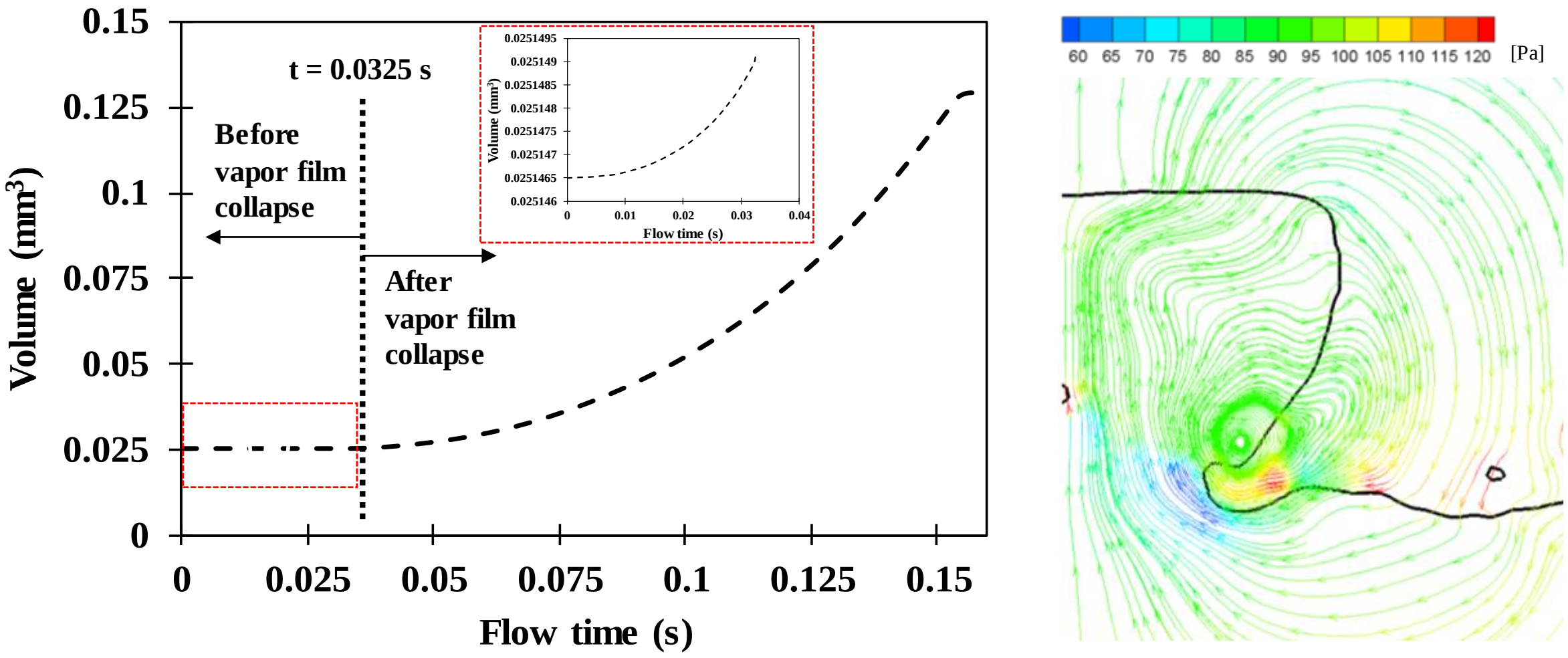

(a) (b)

**Figure 7:** (a) Change of vapor generation rate, before and after the vapour film collapse (b) low pressure zone created in the water phase during rise of vapor bubble

On superheating, the liquefied nitrogen evaporates at a faster rate called homogeneous boiling. The amount of vapor generation before (inset) and after the vapor collapse as expected is very high which is shown in Figure 7(a). Due to homogeneous boiling, micro vapor bubbles are generated in the domain, which was also observed by Liu at al. (2015) in their study. These micro-bubbles either merge together to form a new macro bubble or join with the existing macro vapor bubble (Figure 6(g)). As the vapor bubble rises up, a low pressure zone is created near the water-$N_2$ interface (Figure 7(b)), hence the water just below the vapor bubble also rises along the vapor bubble. Moreover, it also propels the $LN_2$ to fill the void created by rising vapor bubble in the low pressure zone, which pulls the water phase along due to viscous drag. At t = 0.135 s, vapor bubble finally detaches with the water phase and moves towards the outlet, below which smaller bubbles are also rising slowly towards the outlet.

### *3.2.1. Vapor film collapse*

Vapor film collapse occurs whenever the vapor-liquid interface temperature becomes lesser than the Leidenfrost temperature of the boiling fluid. As discussed in the previous section, vapor film collapse leads to direct contact between the boiling fluid and heating fluid. Since the bubble-forming nuclei are absent in case of a liquid surface unlike solid surfaces, the contact leads to superheating of the boiling fluid, which causes rapid change of phase. Hence, predicting the Leidenfrost temperature for various flow conditions is very important. Several studies have been conducted in the past to predict the Leidenfrost point. Berenson (1961) developed a correlation for the minimum temperature difference required to avoid the film collapse in case of on a horizontal solid surface based on the Taylor-Helmholtz hydrodynamic instability, but it was verified with the experimental results of only two fluids, n-pentane and carbon tetrachloride within $\pm 10\%$. Spiegler et al. (1963) derived an expression for Leidenfrost point using Van der Waals equation, which is given as $\frac{27}{32}T_c$, where $T_c$ is the critical temperature. Also it was assumed that the boiling fluid is already superheated to its upper limit i.e. superheat limit temperature ($T_{sl}$) at the initial stage.

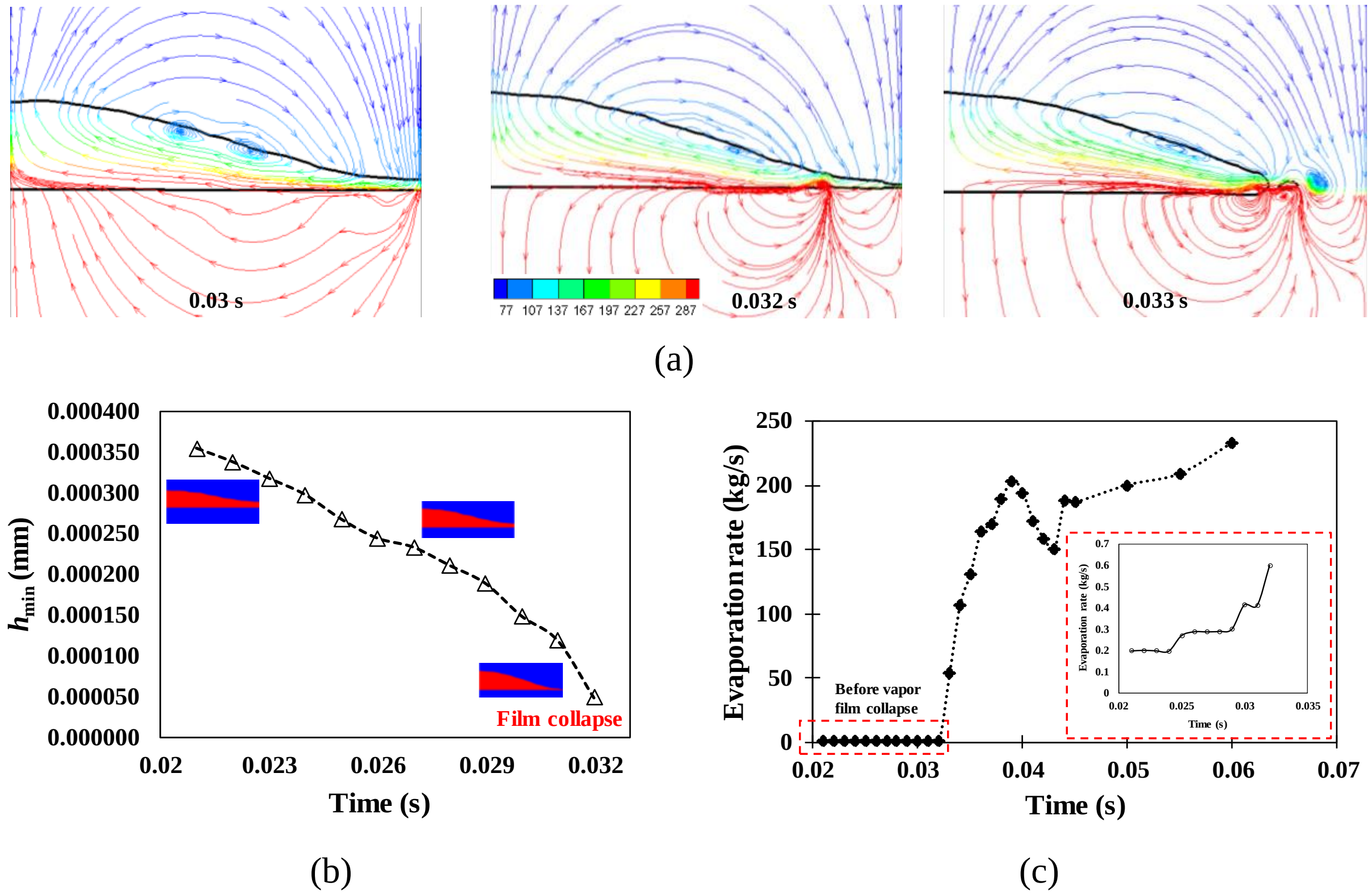

**Figure 8:** (a) Temperature field and streamlines just before and after the vapor film collapse (b) variation of minimum film thickness ($h_{min}$) before rupture (c) change in evaporation rate with time

Baumeister and Simon (1973) developed a method to predict the Leidenfrost point on a horizontal solid surface for cryogens, which included effects of critical temperature of the boiling fluid, thermal properties of the solid surface, surface energy of both, solid and liquid. The correlation given for the Leidenfrost point is $(T_c - T) \times \left[0.16 + 2.4 \times \left(\frac{\rho_l C_{p,l} k_l}{\rho_w C_{p,w} k_w}\right)^{1/4}\right]$, where subscript '$l$' indicates boiling fluid and '$w$' indicates heating fluid, water, whereas, $T_c$, $\rho$, $C_{p,l}$, $k$ are critical temperature, density, specific heat and thermal conductivity, respectively. Recently, Aursand et al. (2018) developed a theoretical model to predict the Leidenfrost point for saturated film boiling on a horizontal surface, where they have used a non-equilibrium evaporation model based on kinetic theory which has also included thermocapillary effects along the evaporating liquid-vapor interface. They have applied a linear stability analysis on the developed model and found that, a film with a very small and a very large film thickness are always unstable, whereas in the case of intermediate film thickness, instability is dependent on the relative dominance of thermocapillary and vapor thrust, where thermocapillary is destabilizing and vapor thrust is stabilizing in nature. They concluded that the mechanism behind the film boiling collapse may be the thermocapillary instability at the liquid-vapor interface.

In the present case of boiling of $LN_2$ on water surface, the vapor film collapses at t = 0.0325 s. Figure 8 shows the temperature field and streamlines for three different instances. At t = 0.032 s, the fluctuation just below the $N_2$ vapor in water phase, indicates the point of inception of film collapse. Reduction of vapor film thickness with time, and its collapse is shown in Figure 8 (b). The minimum thickness just before the rupture at t = 0.323 s is 7.62 $\mu m$, which lies in the region of intermediate scales of film thickness as defined by Aursand et al. (2018). The intermediate film thickness scale lies in the range of 0.1 $\mu m$ < $h_{min}$ < 10 $\mu m$. Since the vapor film collapses in the intermediate film thickness zone, we conclude that the thermocapillary dominates over the vapor thrust. Since the vapor thrust is related to the evaporation rate and the inertia of the vapor mass, and the vapor generation rate is very low in the region (left) before the vapor film collapse as shown in the inset (Figure 8(c)).

## 4. Conclusions

A numerical study of boiling of a cryogen ($LN_2$) on a solid surface, both single-mode and multi-mode, and also on a liquid surface has been computed. A numerical model based on Rayleigh-Taylor instability has been employed to perform the calculations. In case of boiling on a solid surface, bubble frequency, bubble diameter, heat flux increases linearly with increase in wall superheat. Presence of local fluctuations and turbulence since departure of the $1^{st}$ batch of vapor bubbles in case multi-mode boiling on a solid surface, causes asymmetry in bubble growth and departure of $2^{nd}$ batch of vapor bubbles. In the absence of sufficient heat flux, and dominance of thermocapillary over vapor thrust, thickness of initial vapor film decreases with time and collapses at t = 0.0325 s. Vapor film collapse leads to a sharp increase in vapor generation and evaporation rate.

**Acknowledgements**

The authors would like to thank SERB, Department of Science and Technology, Government of India [grant number EMR/2017/000817] for sponsoring this research, and Mr. Lokesh Rohilla for his valuable insights.